\documentclass[aps,prr,superscriptaddress,notitlepage,twocolumn,%bibnotes%,
nolongbibliography
]{revtex4-2}
\usepackage{graphicx,amsmath,mathtools,amsfonts}
\usepackage[english]{babel}
\usepackage{graphicx,amsmath,mathtools,amsfonts}
\usepackage{bm}
\usepackage[normalem]{ulem}
\usepackage[usenames, dvipsnames]{xcolor}
\usepackage{multirow}
\usepackage{tabularx}
\usepackage{soul} % for \st barred text
\usepackage{svg}  
\usepackage{hyperref}
\hypersetup{colorlinks=true, linkcolor=blue, citecolor=blue, urlcolor=blue}
\usepackage{comment}
\usepackage{enumitem}

\usepackage{placeins}

\usepackage{pdfpages}

\makeatletter
\AtBeginDocument{\let\LS@rot\@undefined}
\makeatother

\begin{document}

\title
{
(Anti-)Altermagnetism from Orbital Ordering
\\in the Ruddlesden-Popper Chromates Sr$_{n+1}$Cr$_n$O$_{3n+1}$
}

\author{Quintin N. Meier}
\email{quintin.meier@neel.cnrs.fr}
\affiliation{Univ. Grenoble Alpes, CNRS, Grenoble INP, Institut Néel, 25 Rue des Martyrs, 38042, Grenoble, France}
\author{Alberto Carta}
\affiliation{Materials Theory, ETH Z\"urich, Wolfgang-Pauli-Strasse 27, 8093 Z\"urich, Switzerland}
\author{Claude Ederer}
\affiliation{Materials Theory, ETH Z\"urich, Wolfgang-Pauli-Strasse 27, 8093 Z\"urich, Switzerland}
\author{Andrés Cano}
\affiliation{Univ. Grenoble Alpes, CNRS, Grenoble INP, Institut Néel, 25 Rue des Martyrs, 38042, Grenoble, France}

\begin{abstract}
Altermagnets are collinear antiferromagnets with spin-split electronic states. We introduce Ruddlesden–Popper chromates Sr$_{n+1}$Cr$_n$O$_{3n+1}$ (including SrCrO$_3$) as candidate materials in which altermagnetism can emerge from spontaneous orbital ordering rather than crystal symmetry. First-principles calculations reveal a layer-dependent spin splitting: if the spin and orbital orders align in adjacent layers, the system exhibits non-relativistic spin splitting, and thus altermagnetism. In contrast, if either the spin or the orbital order is reversed in adjacent layers, we observe a layerwise uncompensated spin splitting, that is compensated in the adjacent layer, giving rise to the concept of anti-altermagnetism. In the RP series, odd $n$ members support coexistence of altermagnetism and \textit{anti-altermagnetism}, whereas even $n$ and the perovskite limit are strictly anti-altermagnetic. In both cases, larger $n$ favors metallicity, and in odd $n$ compounds strain can further stabilize altermagnetism.
\end{abstract}

\maketitle
In ferromagnets, the net magnetic moment arises from the splitting and unequal population of spin-up and spin-down electronic states (i.e., time-reversal symmetry breaking). In conventional antiferromagnets, opposite-spin sublattices are related by translations or inversion, so Kramers' theorem enforces full spin degeneracy and effective time-reversal symmetry. Altermagnets form an intermediate case: they exhibit collinear antiferromagnetic order without net magnetization, but the magnetic sublattices are not connected by translations 
or inversion, but by rotations. As a result, the electronic structure shows spin-splitting even without spin-orbit coupling \cite{hayami_momentum-dependent_2019, yuan_giant_2020, ahn_antiferromagnetism_2019, smejkal_crystal_2020, mazin_prediction_2021, ma_multifunctional_2021,smejkal_emerging_2022,mandal_deterministic_2025}. Recent photoemission experiments confirmed the splitting in MnTe \cite{krempasky_altermagnetic_2024, lee_broken_2024}. Beyond its fundamental interest, altermagnetism holds significant potential for antiferromagnetic spintronics \cite{baltz_antiferromagnetic_2018} since it enables ferromagnetic-like responses in compensated magnets  \cite{smejkal_crystal_2020, feng_anomalous_2022, leiviska_anisotropy_2024,gonzalez-hernandez_efficient_2021, bai_observation_2022, surgers_anomalous_2024, badura_observation_2024}. \\

\begin{figure}[b!]
    \centering
    \includegraphics[width=.48\textwidth, clip, trim=4 4 4 4]{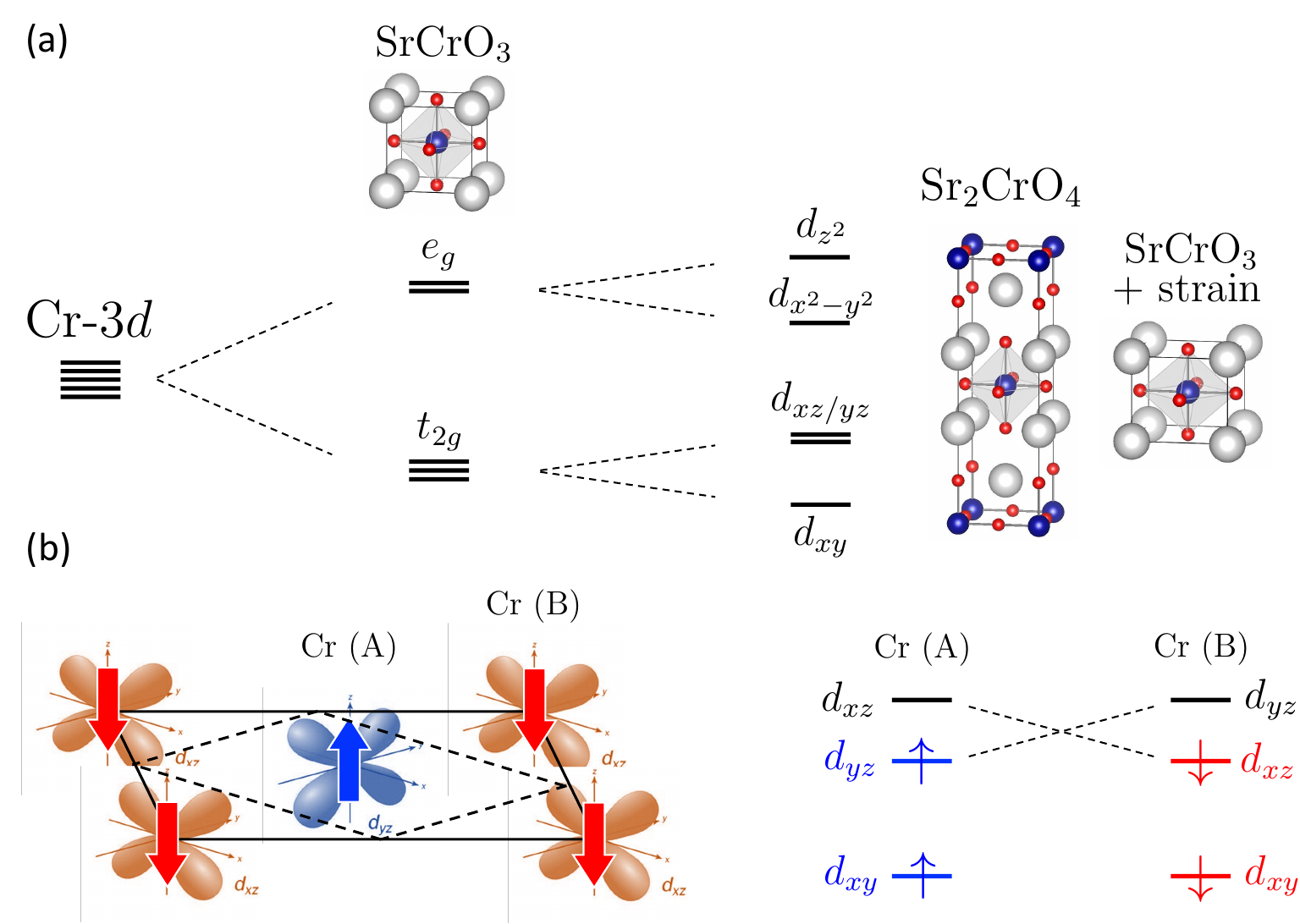}
    \caption{(a) Crystal field splitting of the Cr$^{4+}$ in an octahedral environment results in the splitting into $e_g$ and $t_{2g}$ manifolds. The symmetry breaking of the Ruddelsden-Popper geometry due to the spacers, lowers the energy of the $d_{xy}$ orbitals and splits the $t_{2g}$ manifold into $d_{xy}$ and $d_{xz/yz}$. The same situation can be promoted in the perovskite case via epitaxial strain. (b) The electronic interactions of the  $d_{xz/yz}^1$ manifold further favor the formation of orbital ordering, where either $d_{xz}$ or $d_{yz}$ is occupied on neighboring Cr ions.}
    \label{fig:crystal_field}
\end{figure}

Recently, it was proposed that altermagnetism can arise purely from spontaneous orbital ordering (OO) rather than crystal symmetries \cite{leeb_spontaneous_2024}. However in a bulk material, orbital order tends to either favor ferromagnetism \cite{kugel_jahn-teller_1982}, or the anti-ferromagnetic (AFM) and OO states tend to anti-align according to the Goodenough-Kanamori rules \cite{solovyev_lattice_2006},  which is detrimental for altermagnetism  \cite{leeb_spontaneous_2024}. Here, we propose the series of Ruddlesden-Popper (RP) chromates Sr$_{n+1}$Cr$_n$O$_{3n+1}$ as a new family exemplifying said type of orbital ordering induced altermagnetism.

\begin{figure}[t!]
    \centering  \includegraphics[width=0.45\textwidth, clip, trim=6 6 6 6]{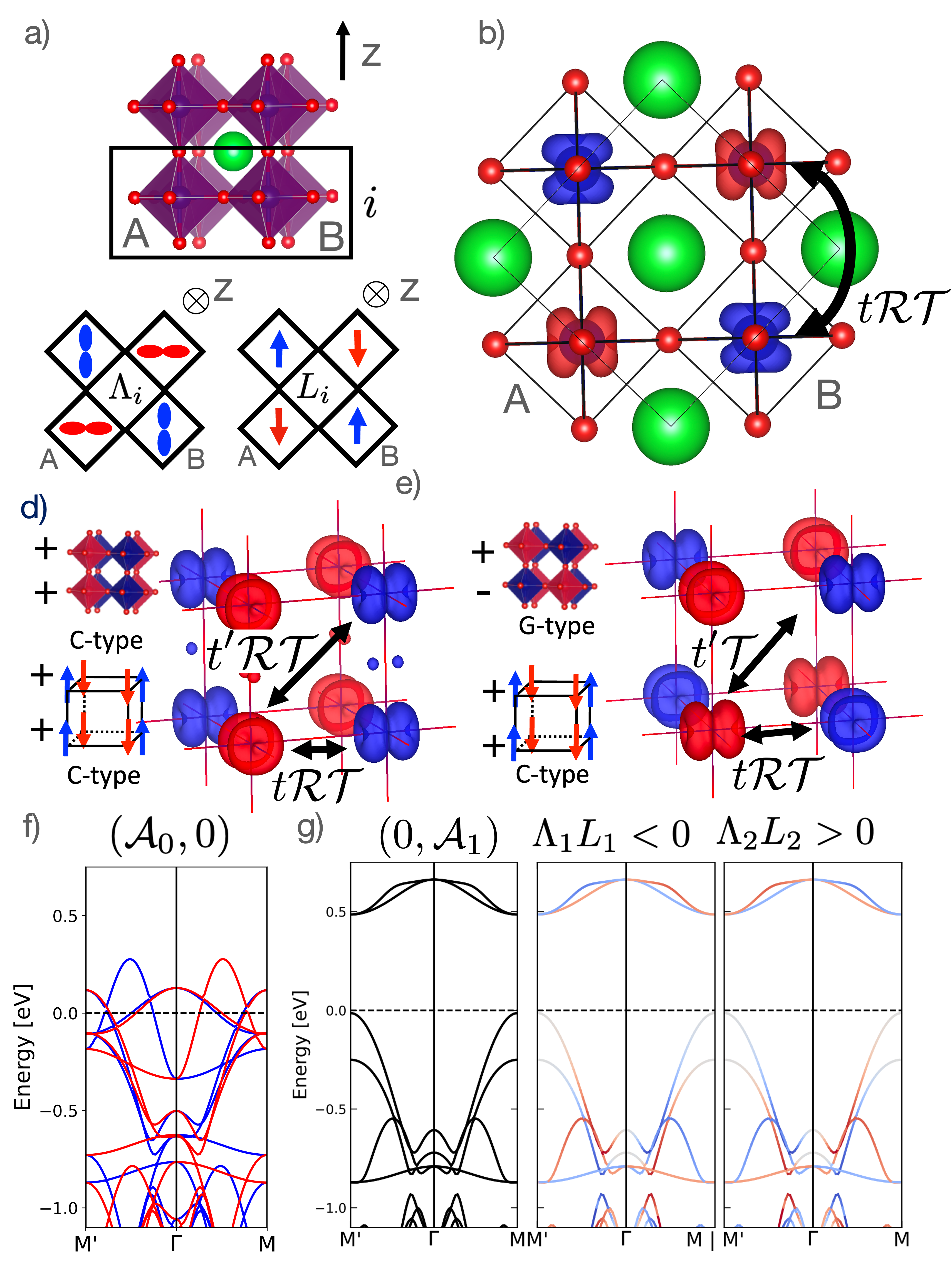}
    \caption{(a) Spin and orbital order parameters of perovskite units (A) and (B) in layer $i$, 
(b) Calculated magnetization density of SrCrO$_3$ ($\rho_\uparrow-\rho_\downarrow$), with red (blue) for spin up (down), showing the $t\mathcal{R}\mathcal{T}$ symmetry connecting magnetic sublattices,
(d) Calculated magnetization density for C-OO/C-AFM and  G-OO/C-AFM order, 
 (f) Calculated  altermagnetic splitting (C-C)  and  (g) anti-altermagnetic (G-C) splitting and projected bands onto each perovskite layer. The isosurfaces are 0.02 $e^-\AA^{-3}$.}
    \label{fig:perovskite}
\end{figure}

 Structurally, the RP Chromates contain $n$ SrCrO$_3$ perovskite layers separated by SrO rocksalt-like spacers [see Fig \ref{fig:crystal_field}(a)]. In all members of the series, the Cr cation exhibits a nominal oxidation state of 4+ ($d^2$ electronic configuration) \footnote{While the possibility of charge disproportionation leading to a mixed valence Cr$^{3+}$ and Cr$^{6+}$ state has been investigated in the case of bulk SrCrO$_3$, this phase was found to be higher in energy compared to phases with Cr$^{4+}$ valence \cite{carta_emergence_2024}. }. As a result, the cubic SrCrO$_3$ perovskite is a rare example of a metallic transition-metal oxide with AFM order \cite{ortega-san-martin_microstrain_2007,arevalo-lopez_electron_2008,komarek_magnetic_2011,zhang_electronic_2015}.
 However, the $d^2$ state in these materials is susceptible to a staggered orbital order arising from a fine-splitting of the octahedral crystal field ($e_g$,$t_{2g}$)  [see Fig. \ref{fig:crystal_field}(a)].   In the Ruddelsden-Popper geometry, an additional splitting of the Cr-$d$ levels leads to an electronic configuration of $d_{xy}^1d_{xz/yz}^1$. In this case, the two-fold degeneracy of the $d_{xz/yz}^1$ orbitals, tends to spontaneously form an orbitally ordered state, in which a staggered $d_{xz/yz}$ occupation is realized [Fig. \ref{fig:crystal_field}(b)]\cite{ishikawa_reversed_2017,aligia_spin_2019,doyle_effects_2024}. This coexistence of OO with collinear AFM induces insulating behavior of the low $n$ members of the series \cite{nozaki_musr_2018,jeanneau_magnetism_2019,jeanneau_singlet_2017, ishikawa_reversed_2017, pandey_origin_2021, doyle_effects_2024}.  Even in the perovskite,  the metallic state is susceptible such an orbital ordered state if a $d_{xy}^1d_{xz/yz}^1$ configuration is realized, for example via epitaxial strain or AFM with electronic correlations \cite{carta_evidence_2022}.  \\
In such a combination of staggered OO and collinear AFM, the OO breaks the translation symmetry between the magnetic sublattices, requiring a rotation of the charge density. Thus, in principle, within each perovskite layer in the RP system fulfills the symmetry requirements of altermagnetism. In a bulk system, the emergence of altermagnetism depends thus on the stacking of OO and AFM in the different layers. To explore this, we define the layer-dependent order parameters 
\begin{equation}
L_i = S_A^i - S_B^i \quad \text{and} \quad \Lambda_i = \Delta n_A^i - \Delta n_B^i,
\end{equation}
where $S_{A(B)}^i$ and $\Delta n_{A(B)}^i$ represent the spin polarization and orbital occupation difference  of the $d_{xz}$ and $d_{yz}$ orbitals ($\Delta n = n_{xz} - n_{yz}$) on the sublattices $A$ and $B$ of layer $i$.  $L_i$ defines the layer-dependent Néel vector, and $\Lambda_i$ is the orbital analogue to the magnetic Néel vector. The resulting OO and AFM order parameters are visualized in Fig. \ref{fig:perovskite}(a). \\

In SrCrO$_3$ ($n=\infty$), the basic building block of the RP structure, density functional theory with Hubbard corrections (DFT+$U$, see Supp. for computational details\cite{supp}) reveal an OO instability if $U$ is large enough (see \cite{carta_evidence_2022} and Supp. Fig. S1 \cite{supp}). The calculated magnetic charge density reveals the OO associated to the occupation of either $d_{xz}$ or $d_{yz}$ on each Cr site in layer $i$, [see Fig. \ref{fig:perovskite}(b)]. (The orbital-dependent filling is quantified is discussed in the end matter~\ref{em2}) While the fully relaxed structure shows a small Jahn-Teller mode accompanying the orbital ordering, the main energy contribution to this distortion is electronic, and the orbital order survives even if the crystal structures show no distortions [See Suppl. Fig. S1 \cite{supp}]. Within each perovskite layer, spin and orbital sublattices coincide, resulting in the orbital ordering breaking the initial translational symmetry between the magnetic sublattices, and thus realizing orbital altermagnetism [Fig. \ref{fig:perovskite}(b)]. 
The simplest stacking of $\Lambda_i$ and $L_i$,  corresponds to C-type AFM and C-type OO, in which $\Lambda_i$ and $L_i$ are aligned in adjacent layers [Fig.~\ref{fig:perovskite}(d)]. In this case, the magnetic and orbital sublattices coincide, and thus enabling $d$-wave altermagnetism \cite{leeb_spontaneous_2024}. This is demonstrated by the clear spin splitting between the $M$ and $\Gamma$ points in the calculated electronic band structure in Fig.~\ref{fig:perovskite}(f).  

However, the ground state of SrCrO$_3$ in the orbitally ordered phase exhibits C-type AFM and G-type OO (see \cite{carta_evidence_2022} and also Suppl. Fig.~S1 \cite{supp}). This situation can be understood in terms of a Kugel-Khomskii model \cite{kugel_jahn-teller_1982, khomskii_orbital_2021}, in which the in-plane exchange is primarily mediated by the $d_{xy}$ orbitals, displaying a typical AFM superexchange mechanism. In contrast, the out-of-plane exchange is mediated by the two-orbital $d_{xz/yz}$ manifold, which favors ferromagnetic and antiferro orbital coupling. Thus, while $L_i$ remains constant in adjacent layers, $\Lambda_i$ changes its sign (see Fig.~\ref{fig:perovskite}(e)). 

\begin{figure}[!htb]
    \centering
    \includegraphics[width=0.9\linewidth]{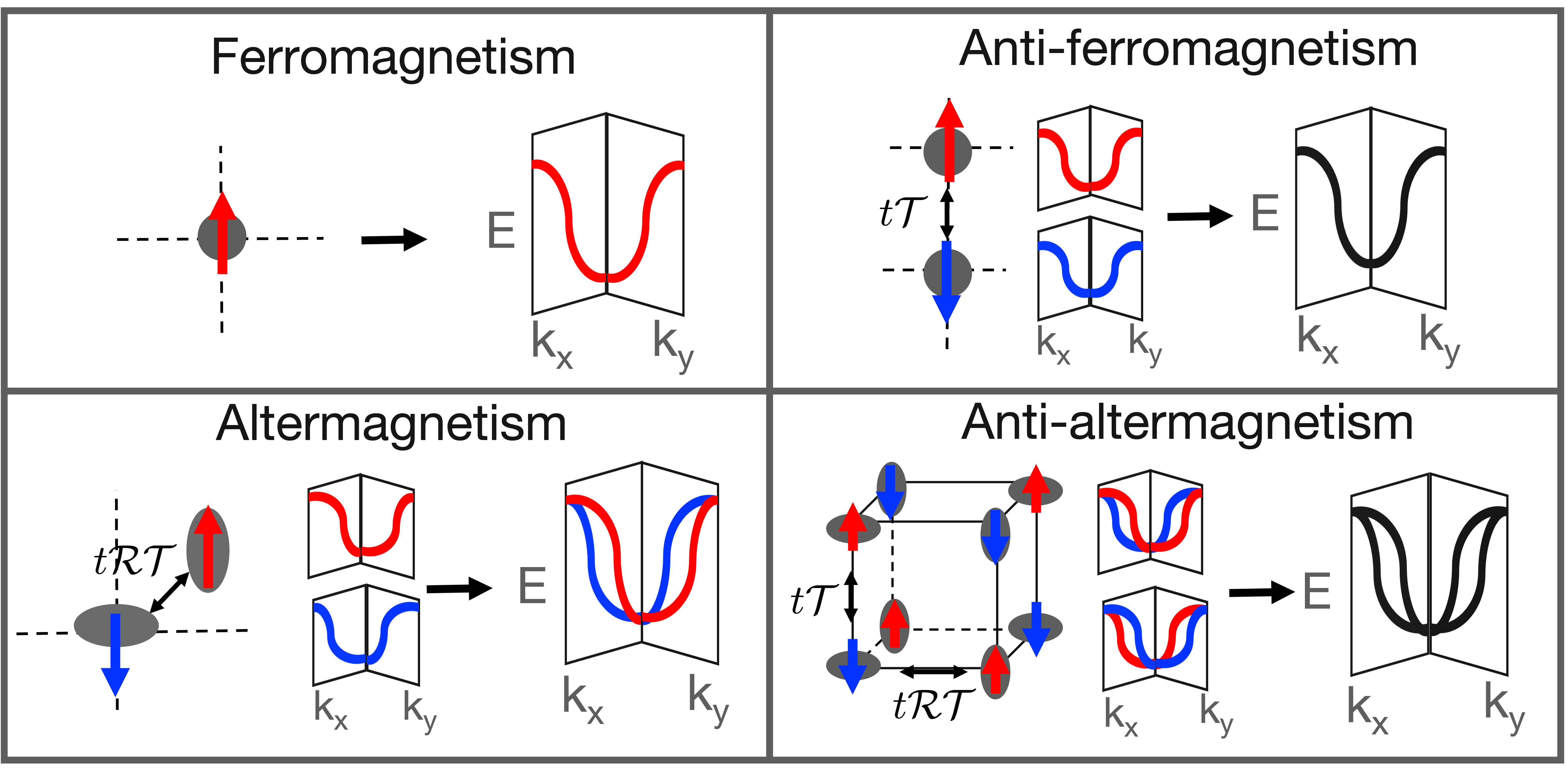}
    \caption{Schematic of magnetic symmetry and band splitting for different magnetic systems: Ferromagnets show a net spin polarization. Antiferromagnets have spin-degenerate bands due to the translational symmetry between sublattices. Altermagnets non-relativistic spin splitting due to combined translation and rotation. Anti-altermagnets host multiple inequivalent sublattice pairs: each displays local altermagnetism, but an additional translation cancels the splitting globally.}
    \label{fig:anti-altermagnetism}
\end{figure}

This situation leads to the concept of anti-altermagnetism. In a conventional antiferromagnet, spin-up and spin-down sublattices are related by time reversal $\mathcal{T}$ combined with a translation $t$ (or inversion), which enforces spin degeneracy. In an altermagnet, the relation instead involves a combined translation and rotation $t\mathcal{R}\mathcal{T}$, allowing compensated non-relativistic spin splitting in parts of the Brillouin zone. Anti-altermagnetism occurs when a material hosts multiple altermagnetic sublattice pairs: within each pair, the spin-up and spin-down sublattices are related only by $t\mathcal{R}\mathcal{T}$ and therefore allowing altermagnetic splitting locally, but the sublattice pairs are connected by a translation $t'\mathcal{T}$ to a sublattice pair of opposite sign, thereby compensating the splitting globally (see Fig.~\ref{fig:anti-altermagnetism}).

In order to classify the different cases we can define the order parameters $(\mathcal A_0 , \bm {\mathcal A})$  based on stacking of the layer-wise orbital and AFM order parameters $\Lambda_i$ and $L_i$. In this case, we define the altermagnetic order parameter $\mathcal{A}_0 = \sum_{j=1}^{N} \Lambda_jL_j$, with $N$ being the total number of layers in the unit cell. Further, a set of possible Néel vectors $\bm{\mathcal{A}}$ can be constructed to describe all the possible anti-altermagnetic cases [End matter\ref{em1}]. This allows us to classify any any collinear AFM material as:\\
$(0,0)$: normal antiferromagnet (AFM); $(\mathcal A _0 , 0)$: altermagnetic (AM);  
$(0, \bm {\mathcal A})$: anti-altermagnet (AAM); $(\mathcal A _0 , \bm {\mathcal A})$: ferri-altermagnet (fAM) \footnote{We note that the the product $\Lambda_i L_i$ transforms as the magnetoelectric octupole $\sim (x^2 - y^2)\mu_z$, where $\mu_z$ is the magnetization density. Thus, the above classification generalizes the description of altermagnetism in terms of magnetoelectric multipoles \cite{bhowal_ferroically_2024}. 
}.\\

In the case of the SrCrO$_3$, 
the relevant order parameters are:    $\mathcal{A}_0 = \Lambda_1L_1 + \Lambda_2L_2,\ 
    \mathcal{A}_1 = \Lambda_1L_1 - \Lambda_2L_2$.
Altermagnetism ($\mathcal{A}_0$) is obtained when the combined orbital/magnetic order is C-type/C-type or G-type/G-type.  while anti-altermagnetism ($\mathcal{A}_1$) is obtained whenever the combined orbital/magnetic order corresponds to either G-type/C-type or C-type/G-type.
 In this way a translational symmetry between the magnetic sublattices of the different layers is recovered [Fig.~\ref{fig:perovskite}(e)]. As a result, the projection of the electronic bands onto Cr atoms in each individual layer reveals a persistent spin splitting, which is compensated in the adjacent layer [Fig.~\ref{fig:perovskite}(g)].  The altermagnetic and anti-altermagnetic character of these configurations are illustrated in Fig~\ref{fig:perovskite}(f,g) and Suppl. Fig~S2 \cite{supp}. The ground state of orbitally ordered SrCrO$_3$ being G-type/C-type orbital/magnetic order thus qualifies it as anti-altermagnetic. \\

Experimentally, such an anti-altermagnetic state could potentially be identified by a spin-splitting of the on the surface or by the identification of the additional splitting of the bulk bands absent in AFM. Further, since the altermagnetic character of each layer is preserved, we can expect altermagnetic phenomena (anomalous Hall responses, spin currents, etc.) within to each layer. The influence on the electronic properties of these effects remain to be explored. \\

This unit-cell dependent splitting suggests that altermagnetism could potentially be engineered from SrCrO$_3$ in ultrathin films or heterostructures. Alternatively, the natural confinement of perovskite layers in the  Ruddelsden-Popper (RP) structure (Sr$_{n+1}$Cr$_{n}$O$_{3n+1}$) may potentially stabilize the altermagnetic phase.  Experimentally, the RP structures are reported in the tetragonal $I4/mmm$ space group \cite{baikie_crystallographic_2007, castillo-martinez_revisiting_2007, jeanneau_structural_2017}.  However, like in the perovskite, the orbital order in the RP structure is to be accompagnied by small JT distortions \cite{doyle_effects_2024} that have not yet been identified experimentally. The relaxed structures obtained from first principles for both the high-symmetry $I4/mmm$ and the JT active phases are listed Supp. Tables I\&II \cite{supp}.\\

\begin{figure}[t!]
    \centering
    \includegraphics[width=0.46\textwidth, clip, trim=6 6 6 6]{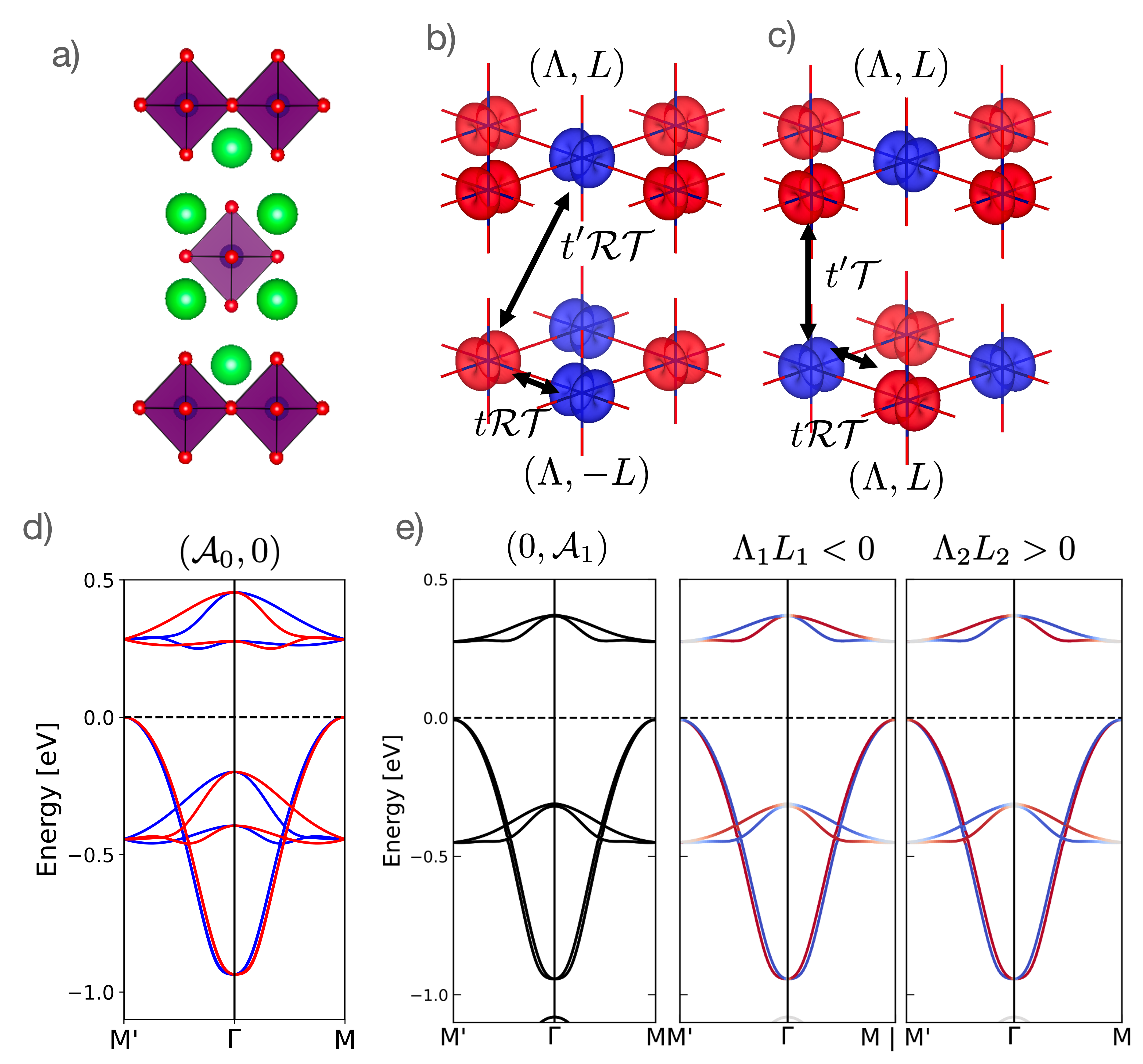}
\caption{(a) Crystal structure of single-layer RP Sr$_2$CrO$_4$. 
(b–c) Isosurfaces (0.02 $e^-\AA^{-3})$ of the magnetic charge density (red = $\uparrow$, blue = $\downarrow$) reveals orbital ordering, yielding either an altermagnetic state (b) or an anti-altermagnetic state (c), where a translation connecting the two layers is recovered. 
This results in altermagnetic band splitting with $\mathcal{A}_0 \neq 0$ (d), or in anti-altermagnetic splitting with $\mathcal{A}_1 \neq 0$ (e), where the total splitting vanishes but projected bands on Cr atoms in different layers still show local spin polarization.}
\label{fig:single_layer}

\end{figure}
\begin{figure*}[htbp!]
    \centering
    \includegraphics[width=0.87\linewidth, clip, trim=5 5 5 5]{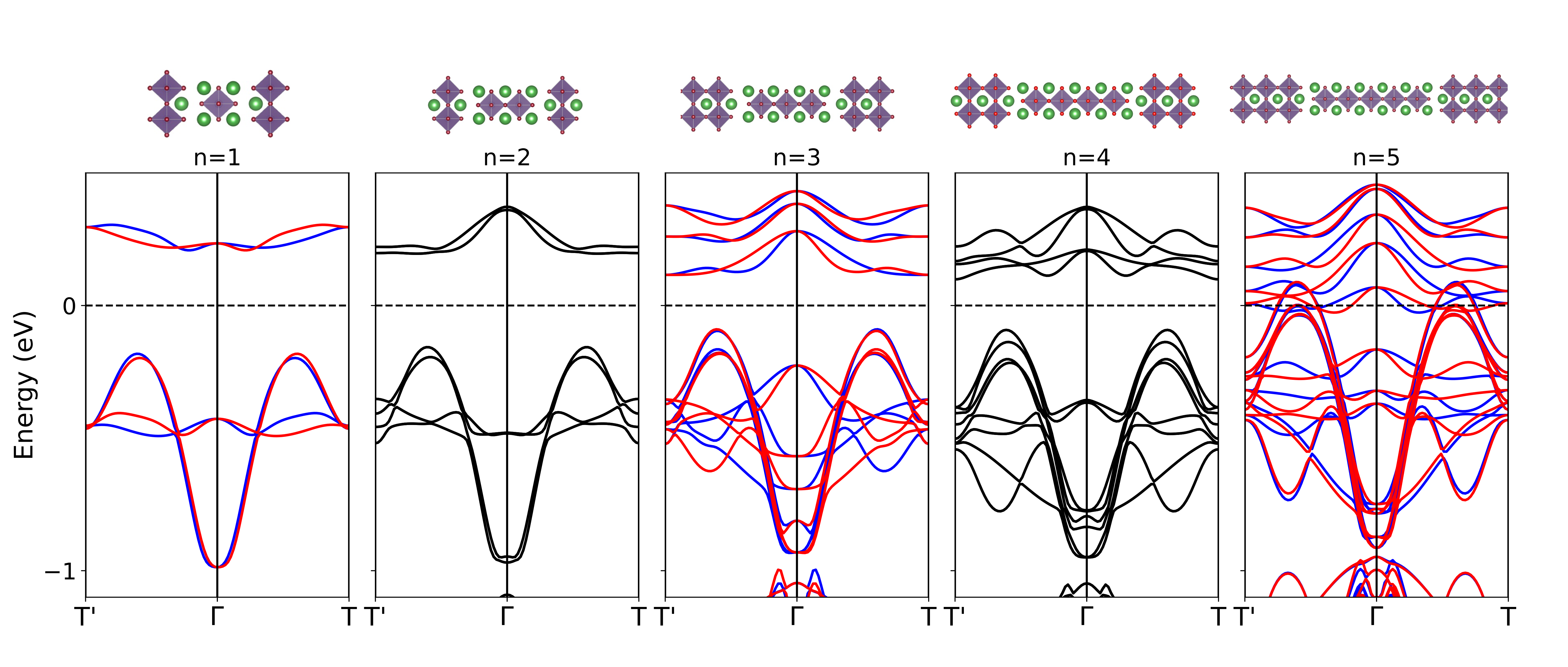}
    \caption{Electronic band structures for different numbers of layers $n$ in Sr$_{n+1}$Cr$_n$O$_{3n+1}$: Compounds with odd numbers of layers $n$ exhibit altermagnetism (red and blue are spin-up and spin-down bands respectively), while compounds with even $n$ show degenerate spin-up and spin-down bands.}
    \label{fig:band_structures}
\end{figure*}
In the single layer Sr$_2$CrO$_4$ ($n=1$) [Fig. \ref{fig:single_layer}(a)],  our calculations reveal two  orbitally ordered states, as we demonstrate using the magnetic charge density Fig~\ref{fig:single_layer}(b-c). (See End Matter \ref{em2} for orbitally projected density of states). To obtain an altermagnetic state ($\mathcal{A}_0 =\Lambda_1L_1+\Lambda_2L_2$), no translational symmetry should connect opposite spins in adjacent layers (e.g. same spin occupies the same orbital in both layers) [Fig.~\ref{fig:single_layer}(b)]\footnote{From symmetry, the layer-wise orbital (AFM) order transforms as $X_2^+$ ($mX_2^+$), yielding four orbital ground states $(\pm \Lambda_1,\pm \Lambda_2)$ and four magnetic ones $(\pm L_1,\pm L_2)$. Their combination produces 16 magneto-orbital domain states}. The resulting magnetic space group is $Cmca'$, and the material displays $d$-wave altermagnetism [Fig.~\ref{fig:single_layer}(d)]. 

As in the perovskite, orbital order in the single-layer RP phase is accompanied by weak JT distortions. However, our calculations reveal that both orbital ordering and altermagnetism persist even when ions remain in the high-symmetry $I4/mmm$ positions (see Suppl. Figs.~S1\&3 \cite{supp}). In fact, the JT distortion has only a minor effect on the altermagnetic splitting of the $d_{xz/yz}$ bands, since these orbitals are either fully occupied or empty irrespective of JT amplitude. Instead, it induces an additional spin splitting of the $d_{xy}$ band, which scales with the distortion amplitude and reflects a more conventional, symmetry-driven mechanism. Thus, the small influence of the JT distortion on the spin splitting establishes orbital ordering as the origin of altermagnetism in these systems.
\\
Likewise, the single layer also supports an anti-altermagnetic state ($\mathcal{A}_1 = \Lambda_1L_1-\Lambda_2L_2$) [Fig.~\ref{fig:single_layer}(c)], with magnetic space group $Pbam'$. In this case, the bands show no net spin splitting globally, but projections onto individual perovskite layers reveals the alternating local splitting [Fig.~\ref{fig:single_layer}(e)]. The energy difference with respect to the altermagnetic state is below 0.1 meV/atom and depends on $U$ and strain [Suppl. Fig.~S1 \cite{supp}].

Due to the effective degeneracy of these two configurations, we expect a mixture of altermagnetic and anti-altermagnetic states in the bulk  material.  However, we identify a coupling between the spin and orbital order parameters and the strain $ \epsilon_{xy}$ given by the invariant
\begin{align}
F =( g_\Lambda \Lambda_1 \Lambda_2 + g_L L_1 L_2) \epsilon_{xy}\quad .
\end{align}
According to our calculations [Suppl. Fig S4 \cite{supp}], the sign of both $g_\Lambda$ and $g_L$ is positive.  Thus, the application of symmetry breaking orthorhombic strain $ \epsilon_{xy}$ offers a potential route to favor the altermagnetic configuration.\\

Finally, we discuss the properties of the other members of the Sr$_{n+1}$Cr$_n$O$_{3n+1}$ series, whos magnetostructural unit cell are composed by shifted blocks of $n$ perovskite layers.
In the ground state of these systems, $\Lambda_i$ changes sign between layers while $L_i$ remains uniform \cite{doyle_effects_2024} (Fulfilling the G-OO/C-AFM configuration within each perovskite block).  Consequently, even-$n$ members always display compensated spin splitting (anti-altermagnetism), whereas odd-$n$ systems cannot fully compensate within each block and thus support (ferri-)altermagnetism. This is shown in the band structures in Fig.~\ref{fig:band_structures}, where  non-relativistic spin splitting is shown for $n=1,3,5$ and anti-altermagnetism for $n=2,4$. [See End Matter \ref{em1} for the order parameters]. Increasing $n$ also enhances conductivity (Fig.~\ref{fig:band_structures}, \cite{doyle_effects_2024}), enabling tuning from insulating to metallic behavior. Since altermagnets are promising for spintronics \cite{smejkal_emerging_2022}, metallicity is a key feature. Our calculations predict that odd-$n$, high-$n$ RP chromates can host both altermagnetism and metallicity provided orbital order persist into the metallic phase. \\

In conclusion, we demonstrated that altermagnetism in Sr$_{n+1}$Cr$_n$O$_{3n+1}$ is induced by orbital order rather than crystal symmetry. In bulk SrCrO$_3$, however, alternating spins and orbitals in adjacent layers can restore some translations, giving rise to the notion of anti-altermagnetism, where the uncompensated splitting in each layer is compensated by the adjacent one. In the RP series, the spacer decouples the layers and allowing  altermagnetic or ferri-altermagnetic states for odd $n$, since spin splitting within each block cannot fully be compensated. The dimensionality further allows to tune the properties: increasing $n$ improves metallicity, enabling a transition from altermagnetic insulator to metal provided orbital order survives into the metallic phase. 
Our findings are also relevant to other orbitally ordered systems such as rare-earth vanadates \cite{khaliullin_spin_2001}, and extend to layered oxides like nickelates \cite{bernardini_ruddlesden-popper_2024} and cuprates \cite{smejkal_emerging_2022, wang_discovery_2024}. We also hope that our work motivates further research into the electronic properties of anti-altermagnets.\\

\textit{Acknowledgements} -- The authors thank A. Ralko for helpful discussions. Computational resources were provided by the GRICAD supercomputing center of Université Grenoble Alpes and GENCI Grant No. 2022-AD01091394. Q.N.M. acknowledges support by the France 2030 government investment plan managed by the French National Research Agency under grant reference PEPR SPIN–MALT(ANR-24-EXSP-0006).

\bibliography{references,bib}

\appendix
\section*{END MATTER}

\subsection{(Anti-)Altermagnetic order parameters}\label{em1}
In order to classify the different cases of altermagnetic and anti-altermagnetic behavior in our materials, we construct a set of global order parameters $(\mathcal A_0 , \bm {\mathcal A})$  based on stacking of the layer-wise orbital and AFM order parameters $\Lambda_i$ and $L_i$. We  find that the quantity $\mathcal{A}_0 $ represents the net altermagnetic component:
\begin{align}
\mathcal{A}_0 
& =  \sum_{j=1}^{N} \Lambda_jL_j,
\end{align}
with $N$ being the total number of layers in the unit cell. For the anti-altermagnetic cases, several Néel vectors are possible, but generally all possible Néel vectors can be constructed using a Fourier basis:
\begin{align}   
\mathcal{A}_k & = 
\sum_{j=1}^{N} 
   e^{-i \tfrac{2\pi}{N}(1-j)k} \Lambda_jL_j
\quad \quad (k = 1,\dots,N-1).\end{align} 

Using the entries of $\mathcal A$, any collinear AFM 
material can thus be further classified as\\

\begin{itemize}
\item 
$(0,0)$: normal antiferromagnet (AFM) 
\item 
$(\mathcal A _0 , 0)$: altermagnet (AM),  
\item 
$(0, \bm {\mathcal A})$: anti-altermagnet (AAM).  
\item 
$(\mathcal A _0 , \bm {\mathcal A})$: ferri-altermagnet (fAM). 
\end{itemize}

\
Let us now briefly develop the relevant order parameters for the RP chromate series Sr$_{n+1}$Cr$_n$O$_{3n+1}$. The structures contain 2 perovskite blocks containing $n$ perovskite layers. For the order parameters,we have to distinguish between odd an even $n$:\\ 
\paragraph*{Even $n$}: For even $n$ the relative order parameters are
\begin{align}
    \mathcal{A}_0 &= \dfrac{1}{n}\sum\limits_{j=1}^{n} \Lambda_jL_j +\dfrac{1}{n}\sum\limits_{j=n+1}^{2n} \Lambda_jL_j \\
    \mathcal{A}_{\pm} &= \dfrac{1}{n}\underbrace{\sum\limits_{j=1}^{n}(-1)^{j}\Lambda_jL_j}_{\text{Block 1}} \pm\dfrac{1}{n}\underbrace{\sum\limits_{j=n+1}^{2n}(-1)^{j}\Lambda_jL_j}_\text{Block 2} 
\end{align}
As an example, in the bilayer Sr$_3$Cr$_2$O$_7$, there are two perovskite blocks containing two perovskite layers each, the possible OO-AFM ($\Lambda_iL_i$) patterns are $\left( +,\,-\;\middle|\; +,\,- \right)$ and  $\left( +,\,- \;\middle|\; -,\,+ \right)$, since within each perovskite block the sign has to alternate. Thus, the system can have one of the two possible Néel vectors $(0,\mathcal{A}_\pm)$, with both configurations qualifying as anti-altermagnetic.\\

\paragraph*{Odd $n$}: For odd $n$, the situation is a bit more tricky. We define the order parameters:
\begin{align}
    \mathcal{A}_0 &= \dfrac{1}{n}\sum\limits_{j=1}^{n} \Lambda_jL_j +\dfrac{1}{n}\sum\limits_{j=n+1}^{2n} \Lambda_jL_j \\
    \mathcal{A}_{1} &= \dfrac{1}{n}\underbrace{\sum\limits_{j=1}^{n}(-1)^{j}\Lambda_jL_j}_{\text{Block 1}} +\dfrac{1}{n}\underbrace{\sum\limits_{j=n+1}^{2n}(-1)^{j}\Lambda_jL_j}_\text{Block 2} \\
    \mathcal{A}_{2} &= \underbrace{\sum\limits_{j=1}^{n}\frac{n(-1)^{\,j} + 1}{\,n\,}\Lambda_jL_j}_{\text{Block 1}} + \underbrace{\sum\limits_{j=n+1}^{2n}\frac{n(-1)^{\,j} + 1}{\,n\,}\Lambda_jL_j}_\text{Block 2} 
\end{align}
In this case, taking the trilayer Sr$_4$Cr$_3$O$_{10}$ as an example, we find the possible OO/AFM  ($\Lambda_iL_i$) patterns are $\left( +,\,-,\,+ \;\middle|\; -,\,+,\,- \right)$ and $\left( +,\,-,\,+ \;\middle|\; +,\,-,\,+ \right)$, where within each block the sign alternates. In the first case, the second layer compensates the imbalance of the first layer, leading to the anti-altermagnetic order $(0,\mathcal{A}_1)$. In the second case, the pattern can be described by the linear combination of $(\mathcal{A}_0,\mathcal{A}_2)$. Thus, having both a net altermagnetic component as well as a anti-altermagnetic pattern, the system qualifies as a ferri-altermagnet.

\subsection{Orbital projected density of states}\label{em2}
In the main text we provide the isosurfaces of the magnetic charge density to reveal the orbital order. Here, we show in addition the orbitally projected DOS showing the staggered, almost complete filling/depletion of the $d_{xz/yz}$ orbitals in both the perovskite and the single layer Sr$_2$CrO$_4$. 
\begin{figure}[htb!]
    \centering
    \includegraphics[width=0.8\linewidth]{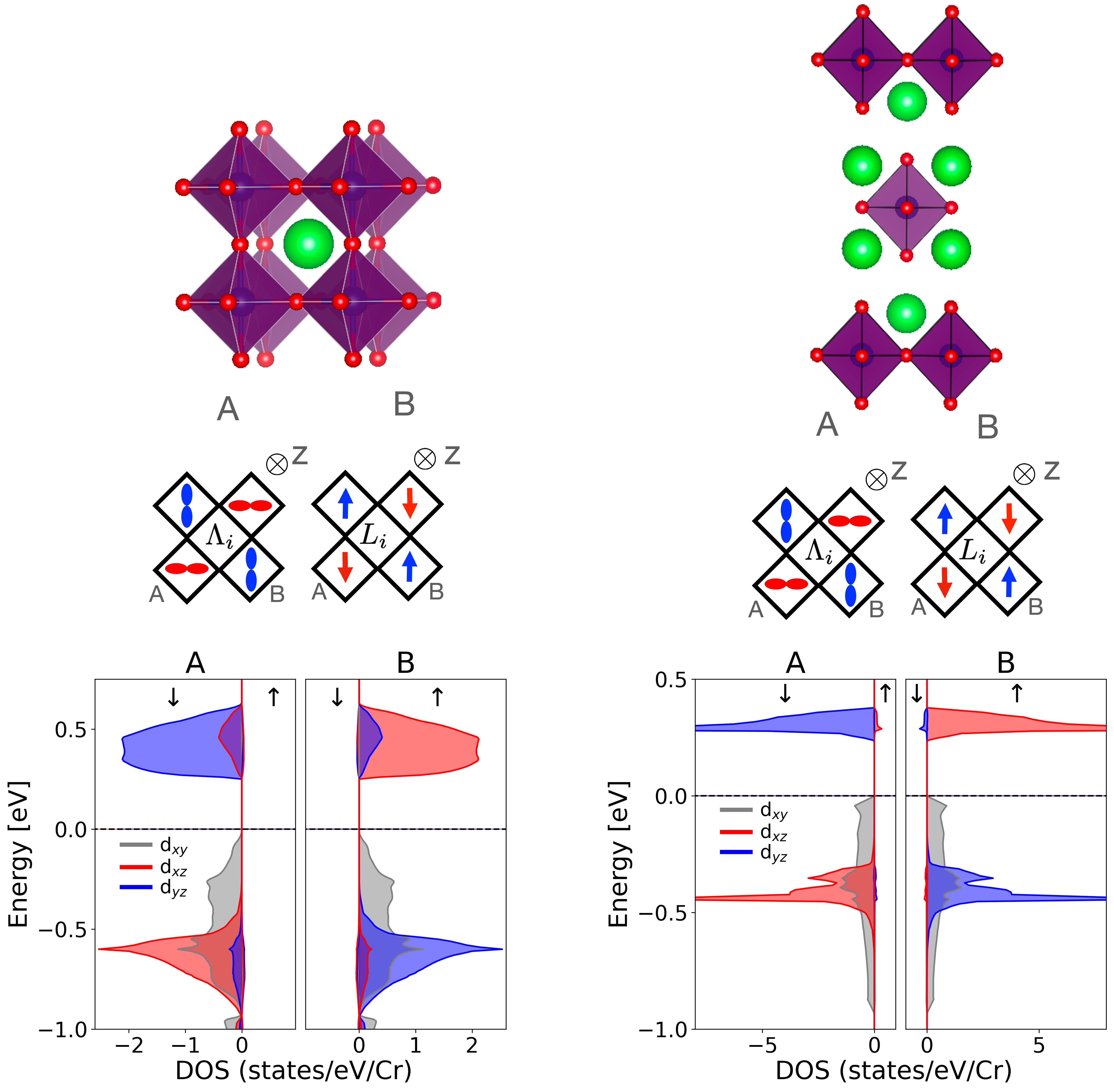}
    \caption{Projected density of states on Cr-$d$ revealing the orbital order of the $d_{xz/yz}$ orbital.}
    
\end{figure}
The single layer RP shows a much reduced bandwidth for the $d_{xz/yz}$ orbitals, which might related to the  orbital ordering being more stable in this case.

% --- end of your main text ---
\clearpage
\onecolumngrid
\clearpage

\begingroup
\pagenumbering{gobble}
\includepdfset{
  pagecommand=\thispagestyle{empty},
  pagebox=cropbox,
  fitpaper=true,
  turn=false,            % stricter than landscape=false
  nup=1x1                % ensure one page per sheet
}

% Force a hard break after page 1
\includepdf[pages=1]{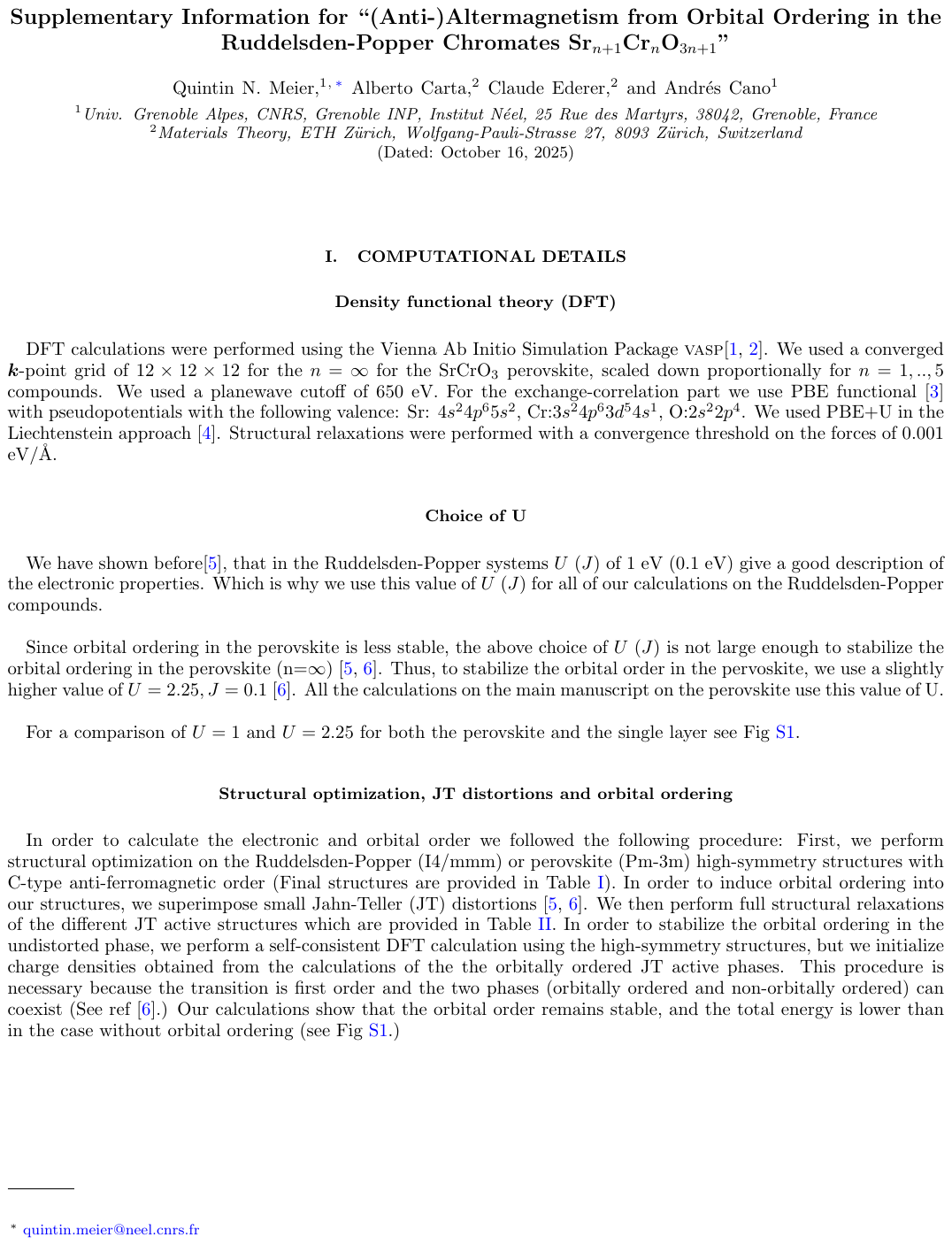}
\clearpage
\includepdf[pages=2]{Supplementary.pdf}
\clearpage
\includepdf[pages=3]{Supplementary.pdf}
\clearpage
\includepdf[pages=4]{Supplementary.pdf}
\clearpage
\includepdf[pages=5]{Supplementary.pdf}
\clearpage
\includepdf[pages=6]{Supplementary.pdf}
\clearpage
\includepdf[pages=7]{Supplementary.pdf}
\end{document}